\documentclass[12pt]{book}
\usepackage{epsfig}
     \makeatletter
     \def\nothing#1{}
     \newdimen\earraycolsep
     \def\eqnarray{\let\@currentlabel\theequation
     \global\@eqnswtrue\m@th
     \global\@eqcnt\z@\tabskip\@centering\let \\\@eqncr
     $$\halign to\displaywidth\bgroup\@eqnsel\hskip\@centering
     $\displaystyle\tabskip\z@{##}$&\global\@eqcnt\@ne
     \hskip 2\earraycolsep \hfil$\displaystyle{##}$\hfil
     &\global\@eqcnt\tw@ \hskip 2\earraycolsep 
     $\displaystyle\tabskip\z@{##}$\hfil
     \tabskip\@centering&\llap{##}\tabskip\z@\cr}
     \renewcommand{\theequation}{\arabic{equation}}
     \renewcommand{\thetable}{\arabic{table}}
     \renewcommand{\thefigure}{\arabic{figure}}

     \def\title{\chapter}
     \renewcommand\chapter{\ifodd\c@page\clearpage\else\cleardoublepage\fi
		    \global\@topnum\z@
		    \@afterindenttrue
		    \secdef\@chapter\@schapter}
     \def\@makechapterhead#1{%
       \vspace*{120\p@}%
       {\parindent \z@ \raggedright \reset@font
         \bfseries #1\par
         \nobreak
         \vskip 36\p@
       }}
     \def\author#1{{\pretolerance=10000 \raggedright \advance \leftskip by 
          1in \noindent #1 \vskip 1pc}}
     \def\affiliation#1{{\advance\leftskip by 1in \noindent #1 \vskip -1pc}}
     \def\refnote#1{{$^{\hbox{\scriptsize #1}}$}}
     \def\affnote#1{\llap{$^{\hbox{\scriptsize #1}}$}}
     
     \renewcommand\section{\@startsection{section}{1}{\z@}{2pc \@plus 1ex minus
         .2ex}{1pc \@plus .2ex}{\reset@font\normalsize\bfseries}}
     \renewcommand\subsection{\@startsection{subsection}{2}{\z@}{1pc \@plus 1ex
         minus.2ex}{1pc \@plus .2ex}{\reset@font\normalsize\bfseries}}
     \renewcommand\subsubsection{\@startsection{subsubsection}{3}{\parindent}
     	{1pc \@plus 1ex minus.2ex}{-0.5em}{\reset@font\normalsize\bfseries}}
     
     \def\AmS{{\protect\the\textfont2 
         A\kern-.1667em\lower.5ex\hbox{M}\kern-.125emS}}

     \def\p@LaTeX{{\family{times}\series{m}\shape{n}\selectfont 
         L\kern-.36em\raise.3ex\hbox{\scriptsize A}\kern-.15em 
         T\kern-.1667em\lower.7ex\hbox{E}\kern-.125emX}}
     
     \newlength{\colwidth}
     
     \def\@oddhead{\hfil}
     \def\@evenhead{\hfil}
     \def\@oddfoot{{\bfseries\hfil\thepage}}
     \def\@evenfoot{{\bfseries\thepage\hfil}}
     \def\fnum@figure{\footnotesize\raggedright{\bfseries 
          \figurename~\thefigure.}}
     \def\fnum@table{\normalsize\raggedright{\bfseries \tablename~\thetable.}}
     \long\def\@makecaption#1#2{\vskip 10\p@ {#1 #2\par}}
     \long\def\@makefntext#1{\setbox0=\hbox{$\m@th^{\@thefnmark}$}
          \noindent\hangindent=\wd0 \box0 #1}
     \def\centerfig#1#2#3#4{\vspace*{#2}\relax
         \centerline{\hbox to#1{\special{#4:#3.#4 x=#1, y=#2}\hfil}}}
     \newbox\@atbox
     \long\def\atable#1#2#3{\begin{table}[tbp]\centering\footnotesize
     \setbox\@atbox\hbox{#2}
     \parbox{\wd\@atbox}{\caption{#1}}\par\smallskip
     #2
     \par\smallskip\parbox{\wd\@atbox}{\raggedright #3}
     \end{table}}
     \long\def\htable#1#2#3{\begin{table}[ht]\centering\footnotesize
     \setbox\@atbox\hbox{#2}
     \parbox{\wd\@atbox}{\caption{#1}}\par\smallskip
     #2
     \par\smallskip\parbox{\wd\@atbox}{\raggedright #3}
     \end{table}} % alternate table definition to permit better placement!!!
     \def\@bibitem{\noindent \hangindent=2pc \hangafter=1}
     \def\thebibliography{%
     \section*{REFERENCES}%
     \bgroup\footnotesize
     \def\newblock{\hskip .11em plus.33em minus.07em}%
     \let\bibitem\@bibitem}
     \def\endthebibliography{\par\egroup}
     \def\@nbibitem#1{\noindent \hangindent=2pc \hangafter=1
     \refstepcounter{enumi}\hbox to 2pc{\arabic{enumi}.\hfil}%
     \immediate\write\@auxout{\string\bibcite{#1}{\arabic{enumi}}}}
     \def\numbibliography{%
     \section*{REFERENCES}%
     \bgroup\footnotesize
     \setcounter{enumi}{0}%
     \def\newblock{\hskip .11em plus.33em minus.07em}%
     \let\bibitem\@nbibitem}
     \def\endnumbibliography{\par\egroup}
     \makeatother
\def\figph#1 #2 #3 #4 {\begin{figure}[ht]
\centerline{\psfig{file=#2.ps,height=#3pt}} \caption[#1]{#4} \label{#1} \end{figure}}

\def\case#1/#2{{\textstyle\frac{#1}{#2}}}
\def\ess{\hskip.444444em plus .499997em minus .037036em}
\def\mss{\hskip.333333em plus .208331em minus .088889em}
\def\sen{\hbox{\scriptsize--}}
\def\eV{e\kern-.10emV }
\def\eVcm{e\kern-.10emV\kern-.15em,\mss}
\def\eVp{e\kern-.10emV\kern-.15em.\ess}
\def\eVc{e\kern-.10emV\kern-.15em/\kern-.10em$c$ }
\def\eVccm{e\kern-.10emV\kern-.15em/\kern-.10em$c$, }
\def\eVcp{e\kern-.10emV\kern-.15em/\kern-.10em$c$. }

\begin{document}

\setcounter{totalnumber}{1}
\setcounter{topnumber}{1}
\setcounter{bottomnumber}{0}
\renewcommand{\topfraction}{1.0}
\renewcommand{\bottomfraction}{0.0}
\renewcommand{\textfraction}{0.0}

\chapter{WHAT INVARIANT ONE-PARTICLE MULTIPLICITY DISTRIBUTIONS AND TWO-PARTICLE
CORRELATIONS ARE TELLING US ABOUT RELATIVISTIC HEAVY-ION COLLISIONS}

\author{\underline{J. Rayford Nix},\refnote{1} Daniel Strottman,\refnote{1}
Hubert W. van Hecke,\refnote{2}\\
Bernd R. Schlei,\refnote{2} John P. Sullivan,\refnote{2} and Michael
J. Murray\refnote{3}}

\affiliation{\affnote{1}Theoretical Division\\
\affnote{2}Physics Division\\
Los Alamos National Laboratory\\
Los Alamos, New Mexico 87545\\
\affnote{3}Cyclotron Institute\\
Texas A\&M University\\
College Station, Texas 77843}

\section{INTRODUCTION}

Many of you are vigorously searching for the quark-gluon plasma---a predicted
new phase of nuclear matter where quarks roam almost freely throughout the
medium instead of being confined to individual nucleons.\refnote{1,2}\ess Such a
plasma is believed to have existed in the first 10~$\mu$s of the universe during
the big bang and could be produced in the laboratory during the little bang of a
relativistic heavy-ion collision.

When nuclei collide head-on at relativistic speeds, the nuclear matter is
initially compressed and excited from normal nuclear density and zero
temperature to some maximum values---during which pions, kaons, and other
particles are produced---and then expands, with a decrease in density and
temperature.  The early stages of the process are often treated in terms of
nuclear fluid dynamics, but at some late stage the expanding matter freezes out
into a collection of noninteracting hadrons.

To sample the density, temperature, collective velocity, size, and other
properties of the system during this freeze-out, some of you are measuring
invariant one-particle multiplicity distributions and two-particle correlations
for the pions, kaons, and other particles that are produced.  Your hope is that
a sharp discontinuity in the value of one or more of the extracted freeze-out
properties as a function of bombarding energy and/or size of the colliding
nuclei could signal the formation of a quark-gluon plasma or other new physics.
For the extraction of these freeze-out properties from your experimental
measurements, a nine-parameter expanding source model was presented at the 12th
Winter Worshop on Nuclear Dynamics.\refnote{3}\ess

\section{NINE-PARAMETER EXPANDING SOURCE MODEL}

This source model describes invariant one-particle multiplicity distributions
and two-particle correlations in nearly central relativistic heavy-ion
collisions in terms of nine parameters, which are necessary and sufficient to
characterize the gross properties of the source during its freeze-out from a
nuclear fluid into a collection of noninteracting, free-streaming
hadrons.\refnote{3\sen 5}\ess The values of these nine parameters, along with
their uncertainties at 99\% confidence limits, are determined by minimizing
$\chi^2$ for the types of data considered.  Several additional physically
relevant quantities, along with their uncertainties at 99\% confidence limits,
can then be directly calculated.  The nine independent source freeze-out
properties that we consider here are the central baryon density $n$, nuclear
temperature $T$, transverse collective velocity $v_{\rm t}$, longitudinal
collective velocity $v_{\ell}$, source velocity $v_{\rm s}$, transverse radius
$R_{\rm t}$, longitudinal proper time $\tau_{\rm f}$, width in proper time
$\Delta\tau$, and pion incoherence fraction $\lambda_{\pi}$.

For a particular type of particle, the invariant one-particle multiplicity
distribution and two-particle correlation function are calculated in terms of a
Wigner distribution function $S(x,p)$, which is the phase-space density on the
freeze-out hypersurface, giving the probability of producing a particle at
spacetime point $x$ with four-momentum $p$.  It includes both a direct
term\refnote{6} and a term corresponding to 10 resonance decays.\refnote{7}\ess
We consider nearly central collisions, assume axial symmetry, and work in
cylindrical coordinates in the source frame, with longitudinal distance denoted
by $z$, transverse distance denoted by $\rho$, and time denoted by $t$.
Throughout the paper we use units in which $\hbar = c = k = 1$, where $\hbar$ is
Planck's constant divided by $2\pi$, $c$ is the speed of light, and $k$ is the
Boltzmann constant.  However, for clarity, we reinsert $c$ in the units of
quantities whose values are given in the text or table.

Integration of the direct term over spacetime leads to the
Cooper-Frye formula for the direct contribution to the invariant one-particle
multiplicity distribution,\refnote{8}\mss namely
\begin{equation}
\label{e:P}
P_{\!\rm dir}(p) = E\,\frac{d^{3\!}N_{\rm dir}}{dp^3} = \frac{1}{2\pi m_{\rm
t}}\,\frac{d^{2\!}N_{\rm dir}}{dy\,dm_{\rm t}} = \frac{2J + 1}{(2\pi)^3}
\int_{\Sigma} d^{3\!}\sigma_{\!\mu} \, \frac{p^{\mu}}{\exp\{[p \cdot v(x) -
\mu(x)]/T(x)\} \mp 1} ~,
\end{equation}
where $E$ denotes the particle's energy, $m_t = \sqrt{m^2 + {p_t}^2}$ its
transverse mass, and $y$ its rapidity.  The quantity $m$ is the particle's rest
mass, and $p_t = \sqrt{{p_x}^2 + {p_y}^2}$ is its transverse momentum.  The
minus sign applies to bosons and the plus sign to fermions.  The quantity $J$ is
the spin of the particle, $v(x)$ is the collective four-velocity, $T(x)$ is the
nuclear temperature, and $\mu(x)$ is the chemical potential for this type of
particle.  We assume that the source is boost invariant within the limited
region between its two ends,\refnote{9,10}\mss and that it starts expanding from
an infinitesimally thin disk at time $t = 0$.  The transverse velocity at any
point on the freeze-out hypersurface, whose integration limits are denoted by
$\Sigma$, is assumed to be linear in the transverse coordinate~$\rho$.

For a particular type of particle, the two-particle correlation function is
given by\refnote{4,5,11}
\begin{equation}
\label{e:C}
C(K,q) = \frac{P_2(p_1,p_2)}{P(p_1)\,P(p_2)} = 1 \pm \lambda\,\frac{|\int
d^{4\!}x\, S(x,K)\, \exp(iq\cdot x)|^2}{[\int d^{4\!}x\, S(x,p_1)] [\int
d^{4\!}x\, S(x,p_2)]} ~,
\end{equation}
where $K = \case 1/2 (p_1 + p_2)$ is one-half the pair four-momentum and $q =
p_1 - p_2$ is the pair four-momentum difference.  In this equation the plus sign
applies to bosons and the minus sign to fermions, and the quantity $\lambda$
specifies the fraction of particles of this type that are produced incoherently.

\section{APPLICATION TO Pb + Pb COLLISIONS AT $p_{\rm lab}/A$ = 158 G\eVc}

We have used the nine-parameter expanding source model described in the previous
section to analyze normalized but still preliminary data from Experiment NA44
performed at CERN's Super Proton Synchrotron.\refnote{12,13}\ess These data
consist of invariant $\pi^+$, $K^+$, and $K^-$ one-particle multiplicity
distributions and $\pi^+$ and $\pi^-$ two-particle correlations for the central
20\% collisions in the reaction Pb~+~Pb at $p_{\rm lab}/A$~=~158~G\eVcp

For symmetric collisions the source velocity $v_{\rm s}$ can be calculated in
terms of the beam momentum per nucleon and nucleon mass, which eliminates the
need to vary this parameter.  The remaining eight adjustable parameters are
determined by minimizing $\chi^2$ with a total of 2137 data points for the five
types of data considered, so the number of degrees of freedom $\nu$ is 2129.
The error for each point is calculated as the square root of the sum of the
squares of its statistical error and its systematic error, with a systematic
error of 15\% for the $\pi^+$, $K^+$, and $K^-$ one-particle multiplicity
distributions and zero for the $\pi^+$ and $\pi^-$ two-particle correlations.
The resulting value of $\chi^2$ is 2165.5, which corresponds to an acceptable
value of $\chi^2/\nu = 1.017$.  There is a 28.6\% probability that $\chi^2/\nu$
would be at least this large for a perfect model.  The individual values of
$\chi^2/\nu$ are 0.904, 0.925, and 0.721 for the $\pi^+$, $K^+$, and $K^-$
one-particle multiplicity distributions and are 1.130 and 1.012 for the $\pi^+$
and $\pi^-$ two-particle correlations, respectively.

The values of the independent freeze-out properties determined this way, along
with their uncertainties at 99\% confidence limits on all quantities considered
jointly, are given in the third column of Table~1.  For comparison, we show in
the second column of Table~1 the corresponding results\refnote{4} for the
central 7\% collisions in the reaction Si~+~Au at $p_{\rm lab}/A$~=~14.6~G\eVc
studied\refnote{14,15} in Experiment E-802 at Brookhaven's Alternating Gradient
Synchrotron.  Compared to these earlier results for Si~+~Au collisions, the
present results indicate that in Pb~+~Pb collisions the freeze-out density is
somewhat lower, the freeze-out temperature is slightly higher, the source at
freeze-out is somewhat larger, and the longitudinal collective velocity is very
poorly determined (because of the limited experimental coverage in rapidity).
The quantity $n_0$ appearing in Table~1 denotes normal nuclear density, whose
value\refnote{16} is calculated from the nuclear radius constant $r_0$ by means
of $n_0 = 3/(4\pi{r_0}^3) = 3/[4\pi(1.16\;{\rm fm})^3] = 0.153\;{\rm fm}^{-3}$.

\htable{Comparison of nine independent source freeze-out properties for the
central 20\% collisions of Pb~+~Pb at $p_{\rm lab}/A$~=~158~G\eVc with those for
the central 7\% collisions of Si~+~Au at $p_{\rm lab}/A$~=~14.6~G\eVcp}
{\vspace{0.8ex}
\begin{tabular}{lccc}
\hline
\vspace{-10pt} \\

& & \multicolumn{2}{c}{\hspace{-10pt}Value and uncertainty at 99\% confidence}
\\

Property & \hspace{12pt} & Si~+~Au & Pb~+~Pb \vspace{2pt} \\
\hline \vspace{-9pt} \\

Central baryon density $n/n_0$ & \hspace{12pt} & 0.145 $^{+0.063}_{-0.045}$ &
0.062 $^{+0.019}_{-0.015}$ \\[2.5pt]

Nuclear temperature $T$ (M\eV) & \hspace{12pt} & 92.9 $\pm$ 4.4 & 95.8 $\pm$ 3.5
\\[2.5pt]

Transverse collective velocity $v_{\rm t}$ ($c$) & \hspace{12pt} & 0.683 $\pm$
0.048 & 0.664 $\pm$ 0.035 \\[2.111pt]

Longitudinal collective velocity $v_{\ell}$ ($c$) & \hspace{12pt} & 0.900
$^{+0.023}_{-0.029}$ & 0.9985 $^{+0.0015}_{-0.94}$
\\[2.111pt]

Source velocity $v_{\rm s}$ ($c$) & \hspace{12pt} & 0.875 $^{+0.015}_{-0.016}$ &
0.9941219 (fixed) \\[2.5pt]

Transverse radius $R_{\rm t}$ (fm) & \hspace{12pt} & 8.0 $\pm$ 1.6 & 11.4 $\pm$
1.5 \\[2.5pt]

Longitudinal proper time $\tau_{\rm f}$ (fm/$c$) & \hspace{12pt} & 8.2 $\pm$ 2.2
& 12.2 $\pm$ 2.1 \\[2.111pt]

Width in proper time $\Delta\tau$ (fm/$c$) & \hspace{12pt} & 5.9
$^{+4.4}_{-2.6}$ & 7.1 $^{+4.5}_{-2.4}$ \\[2.5pt]

Pion incoherence fraction $\lambda_{\pi}$ & \hspace{12pt} & \hspace{0pt} 0.65
$\pm$ 0.11 & \hspace{0pt} 0.690 $\pm$ 0.074
\vspace{2pt} \\
\hline
\end{tabular}}

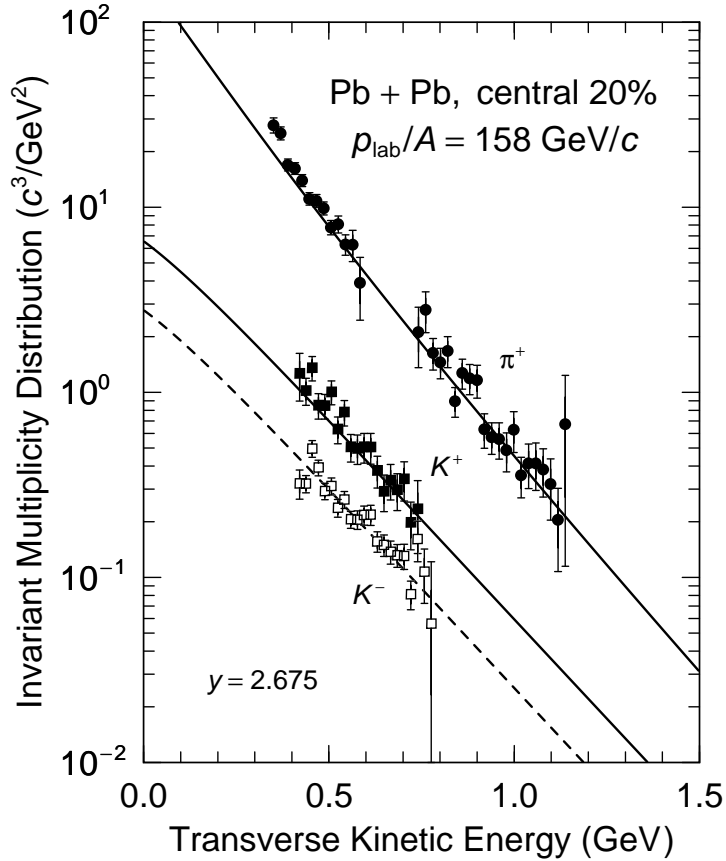
\figph pip fig1 324 {Comparison between model predictions and experimental data
for the invariant one-particle multiplicity distribution $E\,d^{3\!}N/dp^3 =
1/(2\pi m_{\rm t})\,d^{2\!}N/dy\,dm_{\rm t}$ for $\pi^+$, $K^+$, and $K^-$ 
as a function of $m_{\rm t} - m$ for $y = 2.675$.  The error bars shown in this
figure represent statistical errors only.}

Figure~\ref{pip} shows an example of our model predictions, given by the curves,
compared with experimental data for the invariant one-particle multiplicity
distribution for $\pi^+$, $K^+$, and $K^-$.  Solid lines and solid symbols are
used for positive particles, and a dashed line and open symbols are used for
negative particles.  The three curves are calculated for a value of rapidity $y
= 2.675$, which is the central value of the kaon rapidity coverage in this
experiment.  The experimental data for kaons correspond to this same rapidity,
with a rapidity bin width $\Delta y = 0.05$.  For pions, the high-$m_{\rm t}$
experimental data (those to the right of the noticeable break in the
distribution of points) correspond to this same rapidity and bin width, whereas
the low-$m_{\rm t}$ experimental data correspond to the nearby rapidity $y =
2.65$ and rapidity bin width $\Delta y = 0.10$.  The data shown in this figure
represent only a small fraction of that used in our analysis.

Since the two-particle correlation function depends on five variables, it is
somewhat more difficult to graphically compare our model predictions with
experimental data.  The comparisons that we have made thus far involve fixing
the pair rapidity and transverse momentum to specific values and then plotting
the two-particle correlation function versus each component of the pair
three-momentum difference, namely $q_{\rm longitudinal}$, $q_{\rm side}$, and
$q_{\rm out}$, for fixed values of the other two components.  These comparisons
demonstrate that our expanding model satisfactorily reproduces experimental
two-particle correlations.  However, as indicated earlier, the agreement between
model predictions and experimental data is somewhat better for negative pions,
where $\chi^2/\nu = 1.012$, than for positive pions, where $\chi^2/\nu = 1.130$.

\section{RECONCILIATION WITH PREVIOUS ANALYSES}

Analyses with our expanding source model for both the reaction Si~+~Au at
$p_{\rm lab}/A$~=~14.6~G\eVc and the reaction Pb~+~Pb at $p_{\rm
lab}/A$~=~158~G\eVc indicate that the freeze-out temperature is less than 100
M\eV and that both the longitudinal and transverse collective velocities---which
are anti-correlated with the temperature---are substantial.  Similar conclusions
concerning a low freeze-out temperature have also been reached by Cs\"org\H{o}
and L\"orstad.\refnote{17,18}\ess However, other analyses\refnote{12,13,19\sen
21} have yielded a much higher freeze-out temperature of approximately 140 M\eVp
In order to reconcile this serious discrepancy, we now examine the features in
these analyses that led them to the conclusion of a much higher freeze-out
temperature.  These analyses fall into two major classes, which we consider in
turn.

\subsection{Neglect of Relativity in Extrapolation of Slope Parameters}

One type of analysis\refnote{12,13} was based upon the extrapolation to zero
particle mass of extracted slope parameters characterizing the dependence of
unnormalized transverse one-particle multiplicity distributions upon transverse
mass.  For a given reaction and type of particle, this transverse one-particle
multiplicity distribution was represented by the expression\footnote{To
facilitate comparisons with our own expressions, we have transformed the
notation used in Refs.~12~and~13 to that used here.}
\begin{equation}
\label{e:trans}
\frac{1}{m_{\rm t}}\,\frac{dN}{dm_{\rm t}} = A \exp\!\left(-\frac{m_{\rm
t}}{T_{\rm eff}}\right) ~,
\end{equation}
where $A$ is an arbitrary normalization constant and $T_{\rm eff}$ is the
extracted slope parameter.  Values of $T_{\rm eff}$ were extracted in this way
for six types of particles originating from three separate reactions, namely
$\pi^+$, $\pi^-$, $K^+$, $K^-$, $p$, and $\bar{p}$ originating from the reaction
$p$~+~$p$ at center-of-mass energy $\sqrt s$~=~23~GeV\@, from the central 10\%
collisions in the reaction S~+~S at $p_{\rm lab}/A$~=~200~GeV/$c$, and from the
central 6.4\% collisions in the reaction Pb~+~Pb at $p_{\rm
lab}/A$~=~158~GeV/$c$.

As we will see below, the values of these extracted slope parameters contain
valuable information, but they were unfortunately analyzed in Refs.~12~and~13 in
terms of a heuristic equation that neglects relativity, namely
\begin{equation}
\label{e:erron}
T_{\rm eff} = T + m\bar{v}^2 ~,
\end{equation}
where $T$ is the nuclear temperature (whose value we are trying to determine)
and $\bar{v}$ is the average transverse collective velocity of the expanding
matter from which the particle originated.  On the basis of Eq.~(\ref{e:erron}),
the extrapolation in Ref.~13 of the extracted slope parameters to zero particle
mass yielded the result $T$~$\approx$~140~$\pm$~15~M\eVp

In the limit in which the particle velocity is large compared to the average
collective velocity and with the aid of other simplifying assumptions and
approximations, the correct relationship between slope parameter, nuclear
temperature, particle mass, and average collective velocity can be easily
derived from the relativistically correct Eq.~(\ref{e:P}).  With the neglect of
contributions from resonance decays, the neglect of the $\mp 1$ appearing in the
denominator of Eq.~(\ref{e:P}), the assumption of a constant freeze-out
temperature, and the assumption that freeze-out occurs at a constant time $t$ in
the source frame, Eq.~(\ref{e:P}) leads to
\begin{equation}
\label{e:trans2}
\frac{1}{m_{\rm t}}\,\frac{d^{2\!}N}{dy\,dm_{\rm t}} = A' E \int_V d^{3\!}x 
\exp\!\left[-\frac{p \cdot v({\bf x})}{T}\right] = A' E \int_V d^{3\!}x
\exp\!\left\{-\frac{\gamma({\bf x})[E - {\bf p} {\bf \cdot} {\bf v}({\bf
x})]}{T}\right\} ~,
\end{equation}
where $A'$ is a different arbitrary normalization constant from the one
appearing in Eq.~(\ref{e:trans}), the subscript $V$ on the integral denotes the
spatial integration limits for the source, and the position-dependent Lorentz
factor $\gamma({\bf x}) = 1/\sqrt{1 - {\bf v}({\bf x}) {\bf \cdot} {\bf v}({\bf
x})}$.

By introducing an average collective velocity $\bar{v}$ in the integrations in
Eq.~(\ref{e:trans2}), taking the limit in which the particle velocity is large
compared to the collective velocity, specializing to the transverse direction,
and neglecting the pre-exponential $E$ dependence, we are led to
\begin{equation}
\label{e:trans3}
\frac{1}{m_{\rm t}}\,\frac{dN}{dm_{\rm t}} = A
\exp\!\left[-\frac{\bar{\gamma}(m_{\rm t} - p_{\rm t}\bar{v})}{T}\right] = A
\exp\!\left(-\frac{m_{\rm t} - \bar{v}\sqrt{{m_{\rm t}}^2 - m^2}}{T\sqrt{1 -
\bar{v}^2}}\right) ~.
\end{equation}
To obtain the relationship between the slope parameter, nuclear temperature,
particle mass, and average collective velocity, we equate the derivatives with
respect to $m_{\rm t}$ of Eqs.~(\ref{e:trans}) and (\ref{e:trans3}), which leads
to
\begin{equation}
\label{e:T}
T = \left(1 - \frac{\bar{v}m_{\rm t}}{p_{\rm t}}\right) \frac{T_{\rm
eff}}{\sqrt{1 - \bar{v}^2}} = \left(1 - \bar{v}\sqrt{1 + \frac{m^2}{{p_{\rm
t}}^2}}\right) \frac{T_{\rm eff}}{\sqrt{1 - \bar{v}^2}} ~.
\end{equation}
An analogous relationship has also been obtained by Siemens and
Rasmussen\refnote{22} for the case of a blast wave produced by the explosion of
a spherically symmetric fireball.

In the limit of zero particle mass, Eq.~(\ref{e:T}) reduces to
\begin{equation}
\label{e:T2}
T = T_{\rm eff} \sqrt{\frac{1 - \bar{v}}{1 + \bar{v}}} ~,
\end{equation}
which agrees with the result obtained by Schnedermann, Sollfrank, and
Heinz\refnote{23,24} for the case of cylindrical symmetry.  With a typical value
of $0.4\: c$ for the average collective velocity $\bar{v}$ and the limiting
value of $T_{\rm eff}$~$\approx$~140~$\pm$~15~M\eV obtained in Ref.~13 by
extrapolating slope parameters to zero particle mass, Eq.~(\ref{e:T2}) yields
$T$~$\approx$~92~$\pm$~10~M\eV for the nuclear temperature.

\subsection{Several Approximations Made in a Thermal Model}

Another type of analysis\refnote{12,13,19\sen 21} utilized the thermal model of
Schnedermann, Sollfrank, and Heinz\refnote{23,24} to extract the nuclear
temperature and transverse surface collective velocity from unnormalized
experimental transverse one-particle multiplicity distributions.  An
accumulation of effects from several approximations led to a somewhat higher
temperature than we have found with our expanding source model.  These
approximations include the neglect of contributions from resonance decays, the
neglect of the $\mp 1$ appearing in the denominator of Eq.~(\ref{e:P}), the
neglect of the coupling of the transverse motion to the longitudinal motion,
and---most importantly---the neglect of information contained in the absolute
normalization of the multiplicity distributions.  The accumulation of effects
from these approximations was responsible for the conclusion on page 2083 of
Ref.~13 that ``Within a temperature range 100~$\le$~$T$~$\le$~150~M\eVcm the
fits are equally good.''  It is seen that the use of unnormalized experimental
transverse one-particle multiplicity distributions in such a thermal model
provides only a rough indication of the nuclear temperature at freeze-out.

\section{ADDITIONAL STUDIES WITH EXPANDING SOURCE MODEL}

To test the robustness of our expanding source model, we used it to analyze
one-particle and correlation data generated theoretically from a nuclear fluid
dynamical calculation performed with the computer program {\tt
HYLANDER\/}\refnote{25} that corresponded to a high freeze-out temperature and
an extremely low transverse freeze-out velocity.  The results of this study
demonstrated that the model is capable of reproducing the underlying freeze-out
properties even when their values were chosen to lie in unanticipated regions.

To determine whether or not ultrarelativistic proton-proton collisions can be
described in terms of nuclear fluid dynamics, we used our expanding source model
to analyze invariant $\pi^+$, $\pi^-$, $K^+$, $K^-$, $p$, and $\bar{p}$
one-particle multiplicity distributions\refnote{26} for the reaction $p$~+~$p$
at center-of-mass energy $\sqrt s$~=~45~G\eVp We included the systematic and
normalization errors discussed in Ref.~26 in addition to statistical errors.
Because the pion incoherence fraction $\lambda_{\pi}$ does not enter in the
expression for one-particle multiplicity distributions and because for symmetric
collisions the source velocity $v_{\rm s}$ can be calculated, there are only
seven adjustable parameters in this case.  These parameters are determined by
minimizing $\chi^2$ with a total of 459 data points for the six types of data
considered, so the number of degrees of freedom $\nu$ is 452.  The resulting
value of $\chi^2$ is 1132.0, which corresponds to a completely unacceptable
value of $\chi^2/\nu = 2.504$.  The probability that a perfect model would have
resulted in a value of $\chi^2$ at least as large as that found here is the
incredibly small value $4.9 \times 10^{-60}$.

\section{SUMMARY AND CONCLUSIONS}

We have used a nine-parameter expanding source model that includes special
relativity, quantum statistics, resonance decays, and freeze-out on a realistic
hypersurface in spacetime to analyze in detail invariant $\pi^+$, $K^+$, and
$K^-$ one-particle multiplicity distributions and $\pi^+$ and $\pi^-$
two-particle correlations in nearly central collisions of Pb~+~Pb at $p_{\rm
lab}/A$~=~158~G\eVcp These studies confirm an earlier conclusion;\refnote{3\sen
5}\mss for nearly central collisions of Si~+~Au at $p_{\rm lab}/A$~=~14.6~G\eVc
the freeze-out temperature is less than 100 M\eV and both the longitudinal and
transverse collective velocities---which are anti-correlated with the
temperature---are substantial.

We also reconciled our current results with those of previous analyses that
yielded a much higher freeze-out temperature of approximately 140 M\eV for both
Pb~+~Pb collisions at $p_{\rm lab}/A$~=~158~G\eVc and other reactions.  One type
of analysis was based upon the use of a heuristic equation that neglects
relativity to extrapolate slope parameters to zero particle mass.  Another type
of analysis utilized a thermal model in which there was an accumulation of
effects from several approximations.

The future should witness the arrival of much new data on invariant one-particle
multiplicity distributions and two-particle correlations as functions of
bombarding energy and/or size of the colliding nuclei.  The proper analysis of
these data in terms of a realistic model could yield accurate values for the
density, temperature, collective velocity, size, and other properties of the
expanding matter as it freezes out into a collection of noninteracting hadrons.
A sharp discontinuity in the value of one or more of these properties could
conceivably be the long-awaited signal for the formation of a quark-gluon plasma
or other new physics.
  
\section{ACKNOWLEDGMENTS}

We are grateful to Scott Chapman and Arnold J. Sierk for their valuable
contributions during the early phases of this work---especially Scott Chapman's
development of the computer program {\tt freezer\/} that was used in the present
analysis---and to the members of the NA44 Collaboration for their permission
to use their normalized but still preliminary data in the present analysis.
This work was supported by the U.~S. Department of Energy.

\end{document}